\documentclass[twocolumn,showpacs,preprintnumbers,amsmath,amssymb]{revtex4}

\newcommand{\lead}{\mbox{Pb$_{13}$}}

\usepackage{graphicx}
\usepackage{dcolumn}
\usepackage{bm}

\begin{document}


\title{A structural path for the icosahedra$\leftrightarrow$fcc structural transition in clusters}
\author{C. M. Wei}
\affiliation{Institute of Physics, Academia Sinica, Nankang, Taipei 115, Taiwan, R.O.C.}
\author{C. Cheng}%
\affiliation{ Department of Physics, National Cheng Kung
University, Tainan 701, Taiwan, R.O.C.
}%

\author{C. M. Chang}
\affiliation{
Department of Physics, National Dong Hwa University, Hualien 974, Taiwan, R.O.C.
}%

\date{\today}

\begin{abstract}
We propose a structural path for the
icosahedra$\leftrightarrow$fcc transition in clusters and
demonstrate the transition in \lead \ by {\em ab initio}
molecular-dynamics simulation. The proposed path can be described
by using only two variables. The energy surface on this
two-dimensional space for \lead \ was calculated and a barrierless
fcc-to-ico energy path was found. The atomic displacements of the
proposed structural transition for ico and fcc \lead \ were
identified as one of the vibrational eigenmodes of the clusters
with a soft mode for fcc \lead. These agree with the energy
curvatures around the two structures, i.e. the ico \lead \ is at
the bottom of a valley on the energy surface while the fcc \lead \
is at a saddle point. The barriers of this transition for larger
clusters of Pb$_n$ (n=55, 147 and 309) were also calculated, by
{\em ab initio} elastic-band method, and found being smaller than
the room-temperature thermal energy.
%
\end{abstract}

\maketitle

Clusters can have a structure that is prohibited in the
crystallographic translational-symmetry rules, e.g. the icosahedra
(ico) and decahedra (deca) with noncrystalline fivefold symmetry.
Nanoparticles of elements having fcc bulk structure have been
observed to have structures with fivefold symmetry; the most
studied ones include some metallic and rare-gas
clusters\cite{exp1}. As the growth of clusters proceeds beyond a
crossover point, a structural transition to the crystalline
structures
is expected to take place.
Experimentally it has been  observed that structural transitions
in clusters depend on the sizes of clusters as well as
temperature\cite{exp2}. A simple model for the
ico$\leftrightarrow$deca transition, i.e. from one
fivefold-symmetry cluster to another fivefold-symmetry cluster,
has been proposed which involves a cooperative slip
dislocation\cite{exp2,WAu12}. However no microscopic model, to our
knowledge, has ever been proposed for the transitions between the
fivefold-symmetry ico and the crystalline structure of fcc
cuboctahedra (fcc) of the same size. This transition can be a
crucial process near the crossover point from clusters to the
crystalline structure. In this Letter, we present a structural
model for the ico$\leftrightarrow$fcc transition in clusters and
perform {\em ab initio} molecular-dynamics (MD) calculations to
demonstrate that a transition of fcc-to-ico following the proposed
path indeed takes place in \lead.

%

The proposed model of the ico$\leftrightarrow$fcc transition for
13-atom and 55-atom clusters is schematically illustrated in Fig.
1(a) and 1(b). The most left panels show the ico structures with
$I_{h}$ symmetry and the most right panels are the fcc structures
with $O_{h}$ symmetry. Since these structures are highly
symmetrical, all atoms except the center one can be separated into
groups of atoms connected by symmetry operations and thus the
complexity of structural transition under high symmetry constraint
can be greatly reduced. In fact by keeping the cluster within the
$T_h$ symmetry (the common subgroup of $I_h$ and $O_h$), the
cluster can easily transform between the icosahedra ($I_h$
symmetry) and the fcc ($O_h$ symmetry) structures (see Fig. 1).
Specifically for the 13 atom-cluster, it can be viewed as one atom
moving in a two-dimensional (2D) symmetry plane and only two
variables are needed to describe the cluster. The first variable
$r$ is the interatomic distance of the outer atoms to the center
atom. The second variable $s$ describes the angular part
($\Theta=tan^{-1}\sqrt{(2-s)/(2+s)}$) of the outer atom in the 2D
plane. It is easy to verify that if the value of $s$ equals to
$\pm\sqrt{0.8}$, then the structure is ico; and the structure is
fcc if $s$ equals to 0. When the $s$ variable changes from
$\pm\sqrt{0.8}$ to 0, the cluster undergoes the ico-to-fcc
transition as shown in Fig. 1(a).

\begin{figure*}
\includegraphics [width = 2.25 in] {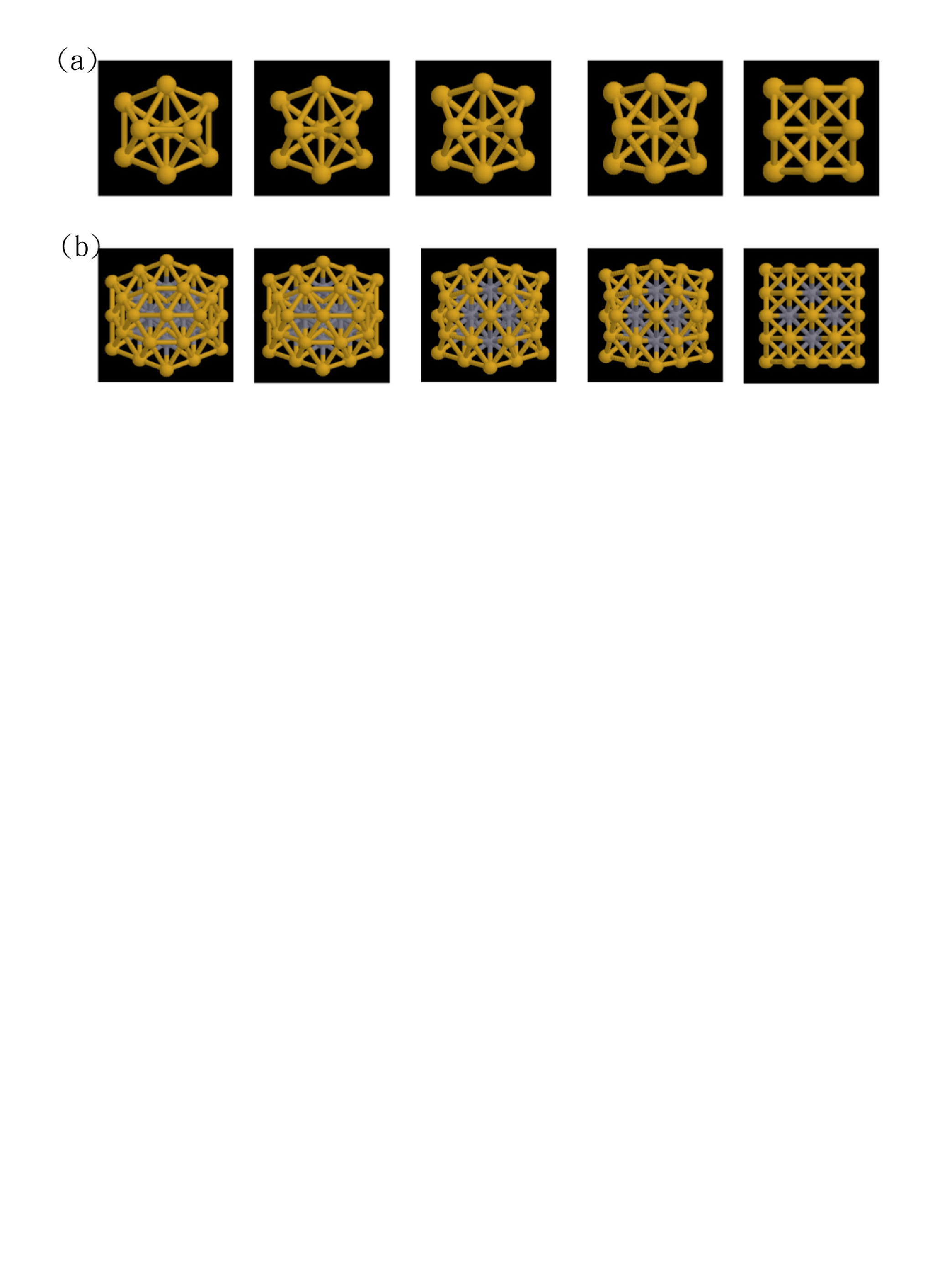}
\caption{\label{fig:wide} The structural transition from ico (the
most left panel) to fcc (the most right panel) in (a) 13-atom and
(b) 55-atom clusters following the change of variable $s$ as
described in the text. }
\end{figure*}


The proposed description of the ico$\leftrightarrow$fcc transition
can be observed in \lead \ by {\em ab initio} MD simulations. All
electronic calculations in this study are based on the density
functional theory\cite{DFT} with the proposed generalized gradient
approximation (GGA) by Perdew, Burke, and Ernzerhof\cite{PBE} for
the non-local correction to a purely local treatment of the
exchange-correlation potential and energy. The single-particle
Kohn-Sham equations\cite{KS} are solved using the plane-wave-based
Vienna {\em ab-initio} simulation program (VASP) developed at the
Institut f\"{u}r Material Physik of the Universit\"{a}t Wien
\cite{VASP}. The interactions between the ions and valence
electrons are described by the projector augmented-wave (PAW)
method\cite{Blochl} in the implementation of Kresse and
Joubert\cite{PAW}. The numbers of treated valence electrons are 4
and 3 for Pb and Al atoms respectively, we shall discuss the Al
clusters later in the Letter. The energy cutoffs for the
plane-wave basis are 100eV and 241eV for Pb and Al systems
respectively. All clusters were simulated by being placed in the
center of cubic supercells which are large enough to neglect
cluster-cluster interactions, i.e. a spacing of at least 10\AA
separated by vacuum between the atoms of neighbouring clusters was
used in the simulations. As the supercells are at least 14\AA  in
length, and going up to 38.4\AA for Pb$_{309}$, only one k point
at the gamma point is included in the Brillouin-zone integration.
Relaxation processes in optimizing static structures are
accomplished by moving atoms to the positions at which all atomic
forces are smaller than 0.02 eV/\AA. The molecular dynamics of the
system was simulated by the canonical Nos\'{e} dynamics\cite{Nose}
with fictitious mass of 2 (in amu) and at least 10000 time steps
of step size 1-femosecond.

%

In the static calculations of \lead, the ico structure was found
more stable than the fcc structure by 2.24 eV (per cluster).
Starting with the fcc structure of \lead \ we performed a few {\em
ab initio} MD simulations of the system at various temperatures.
The initial configurations of the MD simulations were taken as the
static fcc \lead \ plus the randomly set atomic velocities
according to the Maxwell-Boltzmann distribution at the
temperatures of 50K, 100K, 300K and 500K. In all cases the
fcc-to-ico transition, which followed the proposed path we
described, was observed in the beginning of the simulation and
succeeded by thermal vibrations about the ico structure. That the
fcc-to-ico transition takes place in the beginning of the MD
simulations for all the temperatures we studied suggests a
barrierless transition in \lead. To verify this point, the energy
surface of \lead \ on the 2D space of the transition variable $s$
and the interatomic distance $r$ was calculated and the result
presented in Fig. 2. The electronic energy of a total of 169
structures on this 2D space were evaluated, i.e. 13 values of the
variable $s$ ranging from 0.0 to 1.2 and 13 values for the
interatomic distance $r$ ranging from 0.88 to 1.00 (in unit of the
nearest-neighbour distance in bulk Pb) were considered. The ico
\lead \ is at the bottom of the energy surface while the fcc \lead
\ is at a saddle point. According to this 2D energy surface, the
fcc-to-ico transition in \lead \ is indeed barrierless, as implied
in the MD simulation. The energy space of a 13-atom cluster of
course consists of many more degrees of freedom than two. In fact
it consists of 33 degrees of freedom. However, we have
demonstrated by {\em ab initio} MD simulations that the fcc-to-ico
transition in \lead \ does follow the proposed transition path,
though its 33-dimensional energy surface can in principle be very
complicated.

\begin{figure}
\includegraphics [width = 3.0 in]{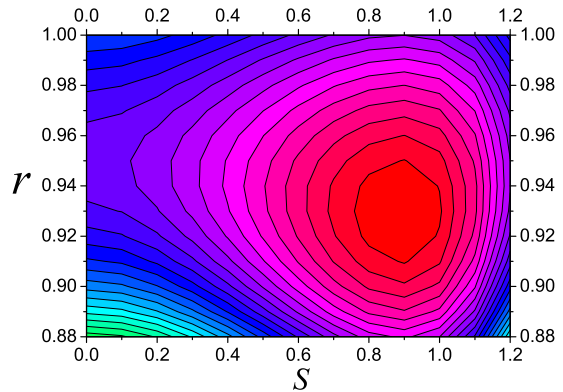}
\caption{\label{fig:epsart}
The calculated energy surface of \lead \ in the 2D space of the variable $s$ and the interatomic distance $r$.
The spacing between the energy contours is 0.2 eV per cluster.
}
\end{figure}


It is expected that, when the transition approaching either
terminal structure, the atomic displacements should be able to be
described by an eigenmode of vibration or a linear combination of
eigenmodes of the terminal structures. We have calculated the
vibrational frequencies of the ico and fcc \lead \ through the
Hessian matrix as provided in VASP. The atomic displacements of
one of the vibrational eigenmodes of the ico and fcc \lead \ were
found to be identical with the proposed description of transition.
The eigenfrequencies are 1.82 (real value) and 1.30 (imaginary
value) THz for the ico and fcc \lead \ respectively. The soft mode
(imaginary frequency) corresponds to the negative curvature of the
energy curve along the transition path at the fcc \lead \ as shown
in Fig. 2. This is the only soft mode found in fcc \lead. That all
the vibrational frequencies obtained for the ico \lead \ are real
indicates ico \lead \ is a stable structure and on the 2D energy
surface it is at the bottom of a valley.

\begin{table}
\caption{\label{tab:table1} The relative energies in the fcc and
ico structures for Pb$_{n}$ clusters and the energy barriers of
the transition through the variable $s$ transition path. The unit
is in eV per cluster except the last line. }
\begin{ruledtabular}
\begin{tabular}{ccccc}
$n$ in Pb$_{n}$  & 13 & 55  & 147  & 309   \\  \hline
ico       & 0.0 & 0.7 & 0.0 & 0.3   \\
fcc       & 2.2 & 0.0 & 3.2 & 0.0   \\
barrier    & 0.0 & 0.1 & 0.3 & 1.5   \\
barrier (eV/Pb)   & 0.0 & 0.0018 & 0.0020 & 0.0049   \\
\end{tabular}
\end{ruledtabular}
\end{table}

This description of ico$\leftrightarrow$fcc transition can in
general be applied to the clusters of larger sizes, for example
see Fig. 1(b) for 55-atom cluster. One thing worth to mention here
is that the proposed transition is a non-diffusive process. As the
fcc-to-ico transition in \lead \ is barrierless, it will be
interesting to know what the barriers are for the larger Pb
clusters. The results of our studies for larger Pb$_n$ clusters,
including n=55, 147 and 309, are summarized in Table I. Firstly
one notices that the relative stability actually oscillates as the
cluster size increases. However, we should make clear that the
comparison here is only for the two structures studied here, i.e.
fcc and ico. Another high-symmetry cluster which is frequently
considered in the studies of clusters is the deca structure. Our
calculations show that the deca structure is the lowest energy
state of the three high-symmetry structures of ico, fcc, and deca
in Pb$_{55}$, the highest in Pb$_{309}$, and in the middle of ico
and fcc in \lead \ and Pb$_{147}$. The barriers for the
ico$\leftrightarrow$fcc transition were obtained from the
elastic-band method with 8 configurations between the terminal
structures of ico and fcc. The initial atomic structures for these
8 configurations were constructed from the structures of the
corresponding variable $s$ as well as the linear scaling of the
bond lengths between the ico and fcc structures. Note that the
barriers described here are the ones from the higher energy state
to the lower one, i.e. it is the barrier of the ico-to-fcc
transition in Pb$_{55}$ and Pb$_{309}$ but the fcc-to-ico
transition in Pb$_{13}$ and Pb$_{147}$. Once a close-shelled
structure, e.g. ico of Pb$_{13}$, is formed, a larger
close-shelled structure, e.g. ico of Pb$_{55}$, can be easily
grown by adding atoms to the smaller one, in spite of the fact
that ico is less stable than fcc in Pb$_{55}$. Then the barrier
needed to be overcome for the transition from ico to the more
stable fcc in Pb$_{55}$ is the barrier we discuss here. According
to Table I, the energy barrier increases as the size of the
cluster increases. We have also listed the energy barriers in
terms of energy per atom which are all considerably smaller than
the classical room-temperature thermal energy. Transitions from
the less stable structure to the more stable one through the
proposed path are therefore likely to occur at low temperature.
The barriers for the reversed processes can be easily evaluated
from adding the energy differences, as provided in the Table I, to
the above discussed barriers. They are either smaller than or in
the order of the room-temperature thermal energy which implies
easy exchange in structures of these clusters at room temperature.

In case the symmetry of the cluster is broken, description of the
proposed transition model requires more than two variables as the
values of $r$ and $s$ are no longer identical for all the atoms
(except the center one) in the cluster. Al$_{13}$ is well-known to
have a distorted ico structure\cite{Al13}. The fcc Al$_{13}$ is
higher in electronic energy than the distorted ico Al$_{13}$ by
0.9 eV (per cluster) in our calculations and the distorted ico is
0.1 eV lower in energy than the undistorted one. We have performed
{\em ab initio} MD calculations for Al$_{13}$, similarly as we did
for \lead, and the transition, roughly following the variable $s$,
was also observed. The proposed path for the
ico$\leftrightarrow$fcc transition is therefore not strictly
confined to the transition between the symmetry group of the ico
and fcc. For clusters of sizes larger than 13 atoms, the number of
variables are more than two even when the symmetry of the cluster
is not broken, e.g. the different interatomic distances between
atoms belonging to different shells. In these systems, the
transition can still possibly follow the proposed path. We have
applied the embeded-atom-method MD to the Pd$_{309}$ clusters and
the fcc-to-ico transition through the proposed path was also
observed.

In conclusion, we have proposed a variable-$s$ description for the
structural transition between the fivefold-symmetry ico structure
and the fcc crystalline structure in clusters and demonstrated,
using {\em ab initio} MD simulation, that in \lead \ the
fcc-to-ico transition did follow this description. We have
calculated the energy surface of the transition in \lead \ which
shows the transition is barrierless. The atomic displacements of
the structural transition in the fcc and ico \lead \ coincide with
one of their vibrational eigenmodes with real and imaginary
frequencies respectively. The barriers of the proposed transition
path for Pb$_n$ (n=55, 147, and 309) were evaluated and found to
be smaller than the room-temperature thermal energy. The current
results have provided a clear-cut microscopic structural model how
the clusters evolve from non-crystalline motifs (ico) to
crystalline structures (fcc).
\\
\\


\begin{acknowledgments}
This work was supported by the National Science Council of Taiwan.
The computer resources were mainly provided by the National Center for High-Performance Computing in HsinChu of Taiwan.
\end{acknowledgments}


\bibliographystyle{unsrt}

\end{document}